\documentclass[twocolumn,pra,showpacs]{revtex4}
\usepackage{amssymb}
\usepackage{amsmath}
\usepackage{graphicx}
\usepackage{amsfonts}

\begin{document}

\title{Coupled cavity QED for coherent control of photon transmission: \\
Green function approach for hybrid systems with two-level doping }
\author{F. M. \surname{Hu}}
\affiliation{$^{1}$ Department of Mathematics, Capital Normal University, Beijing,
100037, China}
\author{Lan \surname{Zhou}}
\affiliation{$^{2}$ Institute of Theoretical Physics, Chinese Academy of Sciences,
Beijing, 100080,China}
\author{Tao \surname{Shi }}
\affiliation{$^{2}$ Institute of Theoretical Physics, Chinese Academy of Sciences,
Beijing, 100080,China}
\author{C. P. \surname{Sun}}
\email{suncp@itp.ac.cn}
\homepage{http://www.itp.ac.cn/~suncp}
\affiliation{$^{2}$ Institute of Theoretical Physics, Chinese Academy of Sciences,
Beijing, 100080,China}

\begin{abstract}
This paper theoretically studies the coherent control of photon transmission
along the coupled resonator optical waveguide (CROW) by doping artificial
atoms in hybrid structures. We provide the several approaches
correspondingly based on the mean field method and spin wave theory et al .
In the present paper we adopt the two-time Green function approach to study
the coherent transmission photon in a CROW with homogeneous couplings, each
cavity of which is doped by a two-level artificial atom. We calculate the
two-time correlation function for photon in the weak-coupling case. Its
poles predict the exact dispersion relation, which results in the group
velocity coherently controlled by the collective excitation of the doping
atoms. We emphasize the role of the population inversion of doping atoms
induced by some polarization mechanism.
\end{abstract}

\maketitle

\section{Introduction}

Recently, all-optical and on-chip setups with the coupled resonator optical
waveguide (CROW) have been implemented experimentally to show the coherent
transmission of \textquotedblleft slow photons" \cite{Fanprl06,nat441},
which is similar to the electromagnetical induced transparency (EIT) effect
\cite{EIT97,EIT01} occurring in the atomic ensemble medium. This successful
experiment implies the coherent couplings of the single mode cavities \cite%
{Fanprl04}, which result in the photonic-crystal-like system with photonic
band structure. For practical applications such as CROW setup they can be
utilized to stop or store the light pulses propagation and then lead to a
quantum device based on these many-body effects, which can also be regarded
as a tunable quantum simulator for the tight binding fermion system in
condensed matter physics.

On the other hand, recently it was discovered \cite{Bqp06159,Pqp06097} that,
when such an array of coupled cavities is doped with two-level atoms, the
photon-blockaded phenomenon can emerge and achieve a Mott insulator state of
polaritons that are many-body dressed states of doped atoms coupled to
quantized modes of optical field in the CROW. Most interestingly, such a
hybrid system with a two-dimensional array of coupled optical cavities in
the photon-blockaded regime will undergo a quantum phase transition from
characteristic Mott insulator (excitations localized on each site) to
superfluid (excitations delocalized across the lattice) \cite{lqpt}. A
similar coplanar hybrid structure based on superconducting circuit has been
proposed for the coherent control of microwave-photons propagating in a
coupled transmission line resonator (CTLR) waveguide. Here, each cavity is
coupled to a tunable charge qubit \cite{sup}. While the CTLR forms an
artificial photonic crystal with an engineered band structure, the charge
qubits collectively behave as spin waves in the low-excitation limit, and
these charge qubits modify the photonic band with energy gaps to slow or
even stop the microwave propagation in this CTLR waveguide. The conceptual
exploration here suggests an electromagnetically controlled quantum device
based on the on-chip circuit for the coherent manipulation of photons, such
as the dynamic appearances of the laser-like output from CTLR waveguide
where the atoms are pumped for some population inversion.

These progressing investigations motivate us to further develop the general
theoretical approach for cavity quantum electrodynamics (QED) with coupled
resonators for coherent manipulations of photon transmission in an
artificial photonic band structure, which can be controlled through some new
mechanisms. In Ref.\cite{zhou} we provide an approach based on the mean
filed method and spin wave theory et al for different hybrid structures,
which consist of the coupled cavity arrays with homogeneous (or
inhomogeneous) couplings and various multi-level-atom doping.

The present paper adopts the two-time Green function approach to study the
coherent transmission of photons in a CROW with homogeneous couplings, each
cavity is doped to a two-level artificial atom. Mathematically the hybrid
system has the same model as that for CTLR waveguide connected to charge
qubits\cite{sup}, but the Green function can work well for the system which
does not satisfy the low-excitation limit, in which we can even obtain exact
solution \cite{sunprl91}. We calculate the two-time retarded Green function
for photons in the weak-coupling case. Its poles predict the exact
dispersion relation, according to which the group velocity can be coherently
controlled by the collective excitation of the doping atoms. We emphasize
the role of the population inversion of the total doping atoms, which is
induced by some polarization or pump mechanism. The dispersion relation
exhibits some exotic features such as the compressed photonic bandwidth.

The paper is organized as followed. In section II we describe our setup of
the photonic band device CROW interacting with doping atoms. Applying the
retarded two-time Green function theory, in section III, we calculate the
eigenfrequencies of the hybrid photon-atom system to characterize the
coherent features of photon transmission. In section IV we study how the
bandwidth and group velocity of photon transmission can be adjusted by
controlling the doped atoms. Then we consider the effects of damping in both
the local mode of cavity and doped atoms. The stable atomic collective
excitations can result in the coherent output of slow photons with some
laser-like properties. In section V, for the phenomenon of slow photons we
study the effective susceptibility of light propagation in the CROW
interacting with doping atoms. In the appendices we give some necessary
details for the Green function calculations and analyze the quasi-spin wave
structure represented by the Green functions for photons and atoms that we
have obtained.

\section{Model for hybrid structure with photonic bands}

We consider a hybrid structure (illustrated in Fig.1 a )- the coupled cavity
array with doping artificial atoms. Here, $N$ optical cavities with
homogeneous and nearest-neighbor couplings form a one-dimensional periodic
structure, which is similar to the fermion system on tight binding lattice.
In practice, there are two ways to implement such CROW. 1. With photonic
crystals, the coupled cavities are built through regularly breaking the
periodicity of photonic crystal. In the photonic bandgap materials, the
cavities are defined by an array (superlattice) of periodic defects in the
periodic modulation. The inter-cavity hopping of photons is due to overlap
between two cavity mode functions. 2. In an electromagnetically controlled
quantum device based on superconducting circuit \cite{sup}, the CROW is
realized by the superconducting waveguide with coupled transmission line
resonators, while the doping systems are implemented by the biased Cooper
pair boxes.

%%%%%%%%%%%%%%%%%%%%%%%%%%%%%%%%%%%%%%%%%%%%%%%%%%%%%%%%%%%%%%%%%%%%%%%%%%%%
\begin{figure}[ptb]
\includegraphics[bb=90 240 520 540, width=8 cm, clip]{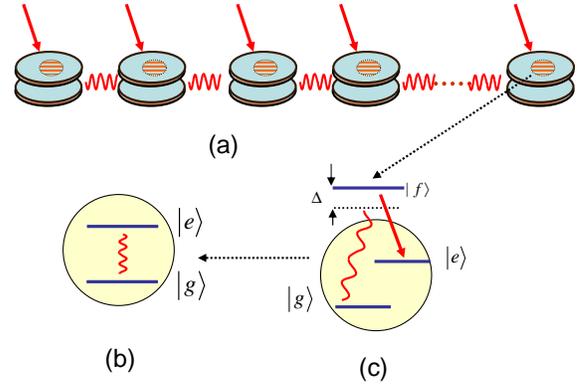}
\caption{(Color online)Configuration of controlled light propagation in a
coupled resonator optical waveguide (CROW) by doping a two-level system (a).
To implement a controllable Rabi transition between the excited and ground
states of the effective two-level system (b), the stimulated Ramann
mechanism is used for a three-level system (c) where the classical
controlling light is resonant between the auxiliary level and the excited
level.}
\label{fig:fig1}
\end{figure}
%%%%%%%%%%%%%%%%%%%%%%%%%%%%%%%%%%%%%%%%%%%%%%%%%%%%%%%%%%%%%%%%%%%%%%%%%%%%%

In Fig.1 (b, c), to implement a Rabi transition with controllable coupling
between the excited state $|e_{\alpha}\rangle $ and ground state $%
|g_{\alpha}\rangle $ of the doping two-level system with level spacing $%
\omega _{A}$, the stimulated Ramann mechanism is usually used for a
three-level system where the classical controlling light is resonant between
the auxiliary level $|f_{\alpha}\rangle$ and the excited level $%
|e_{\alpha}\rangle $.

Actually, for the sake of conceptual simplicity, here we assume that atom in
each cavity only has three energy levels, two metastable lower states $%
|g_{\alpha}\rangle $, $|e_{\alpha}\rangle $ and an auxiliary state $%
|f_{\alpha}\rangle .$ The transition $|f_{\alpha}\rangle \rightarrow
|g_{\alpha}\rangle $ is coupled to a quantized radiation mode with
Rabi-frequency $\Omega ,$ the frequency $\omega _{C}$ and the creation
(annihilation) operator $\hat{a}_{\alpha}^{\dag }$ ($\hat{a}_{\alpha}$) is
in $\alpha th$ cavity, while the transitions $|f_{\alpha}\rangle \rightarrow
|e_{\alpha}\rangle $ are driven by a classical controlling field with
Rabi-frequency $\Omega _{c}$. Moreover, we also assume that the detuning $%
\Delta $ between $|g_{\alpha}\rangle $ and $|f_{\alpha}\rangle $ with
respect to the quantized light is the same as that between $%
|g_{\alpha}\rangle $ and $|e_{\alpha}\rangle $ with respect to the classical
light. Due to the stimulated Ramann effect for large detuning $\Delta $, the
effective coupling can be obtained as $g=\Omega \Omega _{c}^{\ast }/\Delta $%
. In this sense, the effective coupling strength $g$ can be well controlled
by classical Rabi-frequency $\Omega _{c}$ and the detuning $\Delta .$

To describe the collective excitations of the doping atoms, we use the
quasi-spin operators
\begin{align}
\sigma _{\alpha}^{z}& =|e_{\alpha}\rangle \langle
e_{\alpha}|-|g_{\alpha}\rangle \langle g_{\alpha}|\text{,} \\
\sigma _{\alpha}^{+}& =|e_{\alpha}\rangle \langle g_{\alpha}|\text{,}\
\sigma _{\alpha}^{-}=|g_{\alpha}\rangle \langle e_{\alpha}|  \notag
\end{align}%
to express the Hamiltonian $H=H_{A}+H_{AC}+H_{C}$ of the hybrid system.
Here,
\begin{equation}
H_{A}=\sum_{\alpha=0}^{N-1}\frac{\omega _{A}}{2}\sigma _{\alpha}^{z}
\end{equation}%
is the free Hamiltonian of the doping atoms and the interaction between the
local atoms and the corresponding cavity model is of Jaynes-Cummings type
\begin{equation}
H_{AC}=g\sum\limits_{\alpha=0}^{N-1}\hat{a}_{\alpha}\sigma
_{\alpha}^{+}+h.c..
\end{equation}%
It shows a dynamic process that photons are absorbed when atoms transit from
ground state to excited state while the photons are emitted when the atoms
transit from exited state to ground state. The CROW is described by the
Hamiltonian
\begin{equation}
H_{C}=\sum_{\alpha=0}^{N-1}\omega _{C}\hat{a}_{\alpha}^{\dag }\hat{a}%
_{\alpha}+J\sum\limits_{\alpha=0}^{N-1}\hat{a}_{\alpha}^{\dag }\hat{a}%
_{\alpha+1}+h.c.\text{,}
\end{equation}%
where $J$ denotes the inter-cavity coupling. The second term of $H_{C}$
presents the tunneling of photons from the $\alpha th$ cavity to the $%
\alpha+1 th$ one. We notice that the model we adopt above has been used to
demonstrate the photon-blockaded effect most recently\cite{Bqp06159,Pqp06097}%
, and the lasing behavior of the output in line resonators (CTLRs) by
connecting each cavity to a tunable charge qubit in circuit QED\cite{sup}.

To consider the physical significance implied by $H=H_{A}+H_{AC}+H_{C}$, we
perform Fourier transformations
\begin{align}
\sigma _{k}^{+}& =\sum_{\alpha=0}^{N-1}\frac{e^{ik\ell \alpha}}{\sqrt{N}}%
\sigma _{\alpha}^{+},  \notag \\
\sigma ^{z}& =\sum_{\alpha=0}^{N-1}\frac{\sigma _{\alpha}^{z}}{N},
\end{align}%
for \ $k=2\pi n/(\ell N),n=0,1,...,N-1$, with the periodic boundary
condition for the quasi-spin operators $\sigma _{k}^{+},$ $\sigma
_{k}^{-}=(\sigma _{k}^{+})^{\dag }$ and $\sigma ^{z}$. The above Fourier
transformation describes the collective excitations of the spatially
distributed doping atoms as the quasi-spin wave via the collective operators
$\sigma _{k}^{+}$ and $\sigma ^{z}$ \cite{sunprl91}. This is because
\begin{equation}
\sigma _{k}^{+}|G\rangle =\sum_{\alpha=0}^{N-1}\frac{1}{\sqrt{N}}e^{ik\ell
\alpha}|E_{\alpha}\rangle \text{,}
\end{equation}%
represents a spin wave where $|G\rangle =|g_{0}g_{1}g_{2}...g_{N-1}\rangle $
means all atoms are prepared in a ground state, while
\begin{equation*}
|E_{\alpha}\rangle
=|g_{0}g_{1}...g_{\alpha-1}e_{\alpha}g_{\alpha+1}...g_{N-1}\rangle \text{,}
\end{equation*}%
means a single particle excitation in the site $\alpha$.

In order to study the collective excitation described by $\sigma _{k}^{(\pm
)}$ and $\sigma ^{z},$ we consider the corresponding commutation relations
\begin{align}
\left[ \sigma _{k}^{+},\sigma _{k^{\prime }}^{-}\right] & =\sigma
_{kk^{\prime }}^{z},\left[ \sigma _{k}^{+},\sigma _{k}^{-}\right] =\sigma
^{z},  \notag \\
\left[ \sigma ^{z},\sigma _{k}^{+}\right] & =\frac{2}{N}\sigma _{k}^{+},
\notag \\
\left[ \sigma ^{z},\sigma _{k}^{-}\right] & =-\frac{2}{N}\sigma _{k}^{-}%
\text{,}
\end{align}%
where
\begin{equation}
\sigma _{kk^{\prime }}^{z}=\frac{1}{N}\sum_{\alpha=0}^{N-1}e^{-i(k^{\prime
}-k)\ell\alpha}\sigma _{\alpha}^{z}\text{,}
\end{equation}%
means that $\sigma _{00}^{z}=\sigma ^{z}$ , $\sigma _{k}^{+}$ and $\sigma
_{k}^{-}=(\sigma _{k}^{+})^{\dag }$ can not generate a $SU(2)$ subalgebra
except for the case of $k=0.$ Thus $\sigma _{k}^{+}$ and $\sigma _{k^{\prime
}}^{-}$ \ can not be regarded as a collective angular momentum for finite $N$%
.

Applying a discrete Fourier transformation in the $k$-space representation $%
\hat{a}_{k}=\sum_{\alpha}e^{ik\ell \alpha}\hat{a}_{\alpha}/\sqrt{N}$ to the
Hamiltonian $H_{C}$, we have
\begin{eqnarray}
H &=&\sum_{k=0}^{N-1}\Omega _{k}\hat{a}_{k}^{\dagger }\hat{a}_{k}+N\omega
_{A}\frac{\sigma ^{z}}{2}  \label{Model} \\
&&+g\sum_{k=0}^{N-1}(\hat{a}_{k}\sigma _{k}^{+}+h.c.)\text{.}  \notag
\end{eqnarray}%
The photonic band structure is characterized by the dispersion relation
\begin{equation}
\Omega _{k}=\omega _{C}+2J\cos (k\ell )\text{.}
\end{equation}

In principle the above hybrid model can not be solved exactly, but we have
analytically studied its realization based on superconducting circuit \cite%
{sup} when few atoms are populated in their excited state - the
low-excitation limit. In this case this model will be reduced to an exactly
solvable coupling boson model as well as that of all-optical setup for
stopping the light propagating in a CROW in Ref. \cite{Fanprl04}. The
crucial issue of this observation is to use the collective operators \cite%
{sunprl91} $\sigma _{k}^{-}$ and $\ \sigma _{k}^{+}$ as bosonic spin wave
operators in the large $N$ limit with low excitations, since the usual
bosonic commutation relation $\left[ \sigma _{k}^{-},\sigma _{k^{\prime
}}^{+}\right] =\delta _{kk^{\prime }}$ can be approached as $N\rightarrow
\infty $. However, the low excitation requires $\langle \sum_{k}\sigma
_{k}^{+}\sigma _{k}^{-}\rangle \ll N$, which limits the exploitation for the
general cases. So we need to develop a new technique to deal with the
general cases.

\section{Two-time retarded Green function approach for photon transport in
coupled cavity array}

For quantum many-body problem, the quantum or thermal fluctuations near
thermal equilibrium may be characterized by time correlation functions of
the type $\ \left\langle A(t)B(t^{\prime })\right\rangle $, or by the
Fourier transformations of these correlation functions, which give the
correlation fluctuation spectrum. In Heisenberg picture, the time
correlation function $\left\langle A(t)B(t^{\prime })\right\rangle $ of two
observable$\ A$ and $B$ depends only on the time interval $t-t^{\prime }$ by
the invariant of time translation. The propagation of photons in our hybrid
system can be obtained by solving Green function equation for photons.

We consider the linear response of the Green function with respect to an
effective applied driving force, the coupling with doping atoms. We define
the two-time retarded Green function $G_{AB}^{R}\left( t,t^{\prime }\right)
=\langle \langle A(t);B(t^{\prime })\rangle \rangle $ \cite{zub1960} as
\begin{equation}
\langle \langle A(t);B(t^{\prime })\rangle \rangle =-i\theta (t-t^{\prime
})\left\langle \left[ A(t),B(t^{\prime })\right] \right\rangle \text{,}
\end{equation}%
where $\theta (t)=1$ for $t>0$, and $\theta (t)=0$ for $t<0$.

To obtain the equation of motion for Green function, we first list the
equations of motion for the creation and annihilation operators of photons
and atoms by using the Hamiltonian defined in Eq. (\ref{Model}). Then we
have the equations of motion for $G_{AB}^{R}\left( t,t^{\prime }\right) $
\begin{subequations}
\begin{align}
\frac{d}{dt}\langle \langle A(t);B(t^{\prime })\rangle \rangle & =-i\delta
(t-t^{\prime })\langle \lbrack A,B]\rangle  \notag \\
& -i\langle \langle \lbrack A,H];B(t^{\prime })\rangle \rangle , \\
\frac{d}{dt^{\prime }}\langle \langle A(t);B(t^{\prime })\rangle \rangle &
=i\delta (t-t^{\prime })\langle \lbrack A,B]\rangle  \notag \\
& -i\langle \langle A(t);[B,H]\rangle \rangle .
\end{align}%
After applying the Fourier transformation from the time-domain to the
frequency-domain, the retarded Green function is represented as
\end{subequations}
\begin{equation}
\left\langle \left\langle A|B\right\rangle \right\rangle _{\omega
+i\varepsilon }=\int dte^{i(\omega +i\varepsilon )(t-t^{\prime
})}\left\langle \left\langle A(t);B(t^{\prime })\right\rangle \right\rangle ,
\end{equation}%
where $\varepsilon =+0$ is a positive infinitesimal. The equations of motion
for the Green functions can be evaluated in frequency representation

\begin{subequations}
\begin{align}
\omega \langle \langle A|B\rangle \rangle _{\omega }& =\langle \lbrack
A,B]\rangle +\langle \langle \lbrack A,H]|B\rangle \rangle _{\omega }, \\
\omega \langle \langle A|B\rangle \rangle _{\omega }& =\langle \lbrack
A,B]\rangle -\langle \langle A|[B,H]\rangle \rangle _{\omega }.  \label{AB2}
\end{align}%
We notice that in the linear response theory, $\langle \langle A|B\rangle
\rangle _{\omega }$ determines the basic spectrum structure of the hybrid
system through the poles of $\langle \langle A|B\rangle \rangle _{\omega }$
- the physical eigenfrequencies.

With the notations above, we write down the equation of the Green functions $%
\langle \langle \hat{a}_{k}|\hat{a}_{k}^{\dag }\rangle \rangle _{\omega }$, $%
\langle \langle \sigma _{k^{\prime }}^{-}|\sigma _{k}^{+}\rangle \rangle
_{\omega }$, $\langle \langle \sigma _{k}^{-}|\sigma _{k}^{+}\rangle \rangle
_{\omega }$, $\langle \langle \hat{a}_{k}|\sigma _{k^{\prime }}^{+}\rangle
\rangle _{\omega }$, and $\langle \langle \sigma _{k}^{-}|\hat{a}_{k^{\prime
}}^{\dag }\rangle \rangle _{\omega }$. The photon correlation $\langle
\langle \hat{a}_{k}|\hat{a}_{k}^{\dag }\rangle \rangle _{\omega }$ can be
obtained by using the commutation relation between $\hat{a}_{k}$ and $H$ as

\end{subequations}
\begin{equation}
(\omega -\Omega _{k})\langle \langle \hat{a}_{k}|\hat{a}_{k}^{\dag }\rangle
\rangle _{\omega }=1+g\langle \langle \sigma _{k}^{-}|\hat{a}_{k}^{\dag
}\rangle \rangle _{\omega }.
\end{equation}%
We can also calculate the Green function$\ $of the many-atom correlation $%
\langle \langle \sigma _{k^{\prime }}^{-}|\sigma _{k}^{+}\rangle \rangle
_{\omega }$, which satisfies

\begin{align}
\omega \langle \langle \sigma _{k^{\prime }}^{-}|\sigma _{k}^{+}\rangle
\rangle _{\omega }& =-\frac{g}{N}\sum_{k^{\prime \prime
}\alpha}e^{-i(k^{\prime }-k^{\prime \prime })\ell\alpha}\langle \langle
\sigma _{\alpha}^{z}\hat{a}_{k^{\prime \prime }}|\sigma _{k}^{+}\rangle
\rangle _{\omega } \\
-& \frac{1}{N}\sum_{\alpha=0}^{N-1}e^{-i(k^{\prime }-k)\ell\alpha}\langle
\sigma _{\alpha}^{z}\rangle +\omega _{A}\langle \langle \sigma _{k^{\prime
}}^{-}|\sigma _{k}^{+}\rangle \rangle _{\omega }.  \notag
\end{align}

To cut off the Green function hierarchy, we make a \emph{mean field
approximation}\cite{wolff,Kubo}
\begin{equation}
\langle \langle \sigma _{\alpha}^{z}\hat{a}_{k^{\prime \prime }}|\sigma
_{k^{\prime }}^{+}\rangle \rangle _{\omega }\approx \langle \sigma
_{\alpha}^{z}\rangle \langle \langle \hat{a}_{k^{\prime \prime }}|\sigma
_{k^{\prime }}^{+}\rangle \rangle _{\omega }\text{,}
\end{equation}%
where the factor $\langle \sigma _{\alpha}^{z}\rangle $ represents the large
atomic population inversion in the initial state and the light-atom
interaction hardly changes this population. The system of equations of Green
functions\ has an approximately closed form with three simplified equations
(For the details please see the appendix A),
\begin{align}
\langle \langle \hat{a}_{k}|\hat{a}_{k}^{\dag }\rangle \rangle _{\omega }& =%
\frac{1}{\omega -\Omega _{k}}+\frac{F(k,k)}{\omega -\Omega _{k}}\text{,}
\notag  \label{eq:gf} \\
\langle \langle \sigma _{k^{\prime }}^{-}|\sigma _{k}^{+}\rangle \rangle
_{\omega }& =-\frac{\langle \sigma _{kk^{\prime }}^{z}\rangle \lbrack
1+F(k,k)]}{f_{k^{\prime }}\left( \omega \right) }  \notag \\
& -\sum_{k^{\prime \prime }\neq k,\neq k^{\prime }}\frac{\langle \sigma
_{k^{\prime \prime }k^{\prime }}^{z}\rangle F(k^{\prime \prime },k)}{%
f_{k^{\prime }}\left( \omega \right) }\text{,} \\
\langle \langle \sigma _{k}^{-}|\sigma _{k}^{+}\rangle \rangle _{\omega }& =-%
\frac{\langle \sigma ^{z}\rangle }{f_{k}\left( \omega \right) }%
-\sum_{k^{\prime }\neq k}\frac{\langle \sigma _{k^{\prime }k}^{z}\rangle
F(k^{\prime },k)}{f_{k}\left( \omega \right) }\text{,}  \notag
\end{align}%
where we have defined
\begin{align}
F(k,k^{\prime })& =\frac{g^{2}\langle \langle \sigma _{k}^{-}|\sigma
_{k^{\prime }}^{+}\rangle \rangle _{\omega }}{\omega -\Omega _{k}}, \\
f_{k}\left( \omega \right) & =\omega -\omega _{A}+\frac{g^{2}\langle \sigma
^{z}\rangle }{\omega -\Omega _{k}},  \notag \\
\langle \sigma _{kk^{\prime }}^{z}\rangle & =\frac{1}{N}\sum_{%
\alpha=0}^{N-1}e^{-i(k^{\prime }-k)\ell\alpha}\langle \sigma
_{\alpha}^{z}\rangle .  \notag
\end{align}

\begin{figure}[ptb]
\includegraphics[bb=80 270 520 570, width=6 cm, clip]{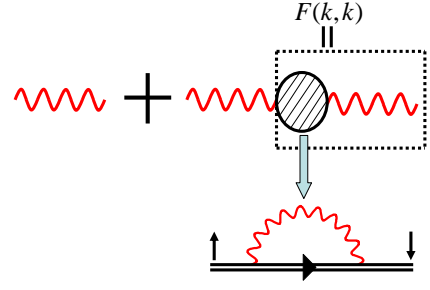}
\caption{The Feynman diagram for the effective photon transmission through
the CROW in a propagating mode. The photon propagator contains the free part
(the bare photon propagator is denoted by a single wiggly line) and the
second perturbation part (one wiggly line plus a box). The box includes the
shade circle for the self energy of photon and the bare atom propagator( the
double arrowed line).}
\label{fig:fig2}
\end{figure}

To consider the basic processes of the photon distribution in k-space, we
draw the Feynman diagram (Fig. \ref{fig:fig2}) to interpret the above
equation of $\langle \langle \hat{a}_{k}|\hat{a}_{k}^{\dag }\rangle \rangle
_{\omega }$ in terms of the basic processes. Here, the photon propagator $%
1/(\omega -\Omega _{k})$ (denoted by a wiggly line) appears twice. In the
second term of the right hand side of first equation in Eq. (\ref{eq:gf}),
it is modified by an interaction with atomic flips characterized by bare
atom propagator $\langle \langle \sigma _{k}^{-}|\sigma _{k^{\prime
}}^{+}\rangle \rangle _{\omega }$, which is denoted by a double line. This
Feynman diagram describes a second order process of the interaction between
the localized modes of the optical field and the doping atoms.
%%%%%%%%%%%%%%%%%%%%%%%%%%%%%%%%%%%%%%%%%%%%%%%%%%%%%%%%%%%%%%%%%%%%%%%%%%%%%
%%%%%%%%%%%%%%%%%%%%%%%%%%%%%%%%%%%%%%%%%%%%%%%%%%%%%%%%%%%%%%%%%%%%%%%%%%%%%

We consider the weak-coupling case. On the right side of the equation of $%
\langle \langle \sigma _{k}^{-}|\sigma _{k}^{+}\rangle \rangle _{\omega }$
in Eq. (\ref{eq:gf}) there are two terms, one is about the zero order of
coupling constant $g$ and the other is about the second or higher order of $g
$. Here we solve the equation of $\langle \langle \sigma _{k}^{-}|\sigma
_{k}^{+}\rangle \rangle _{\omega }$ with the lowest order term of $g$:
\begin{equation}
\langle \langle \sigma _{k}^{-}|\sigma _{k}^{+}\rangle \rangle _{\omega } =-
\frac{\langle \sigma ^{z}\rangle }{f_{k}\left( \omega \right) }\text{.}
\end{equation}%
So we obtain the lowest order solution for
\begin{equation}
\langle \langle \hat{a}_{k}|\hat{a}_{k}^{\dag }\rangle \rangle _{\omega }=%
\frac{\omega -\omega _{A}}{(\omega -\omega _{A})(\omega -\Omega
_{k})+g^{2}\left\langle \sigma ^{z}\right\rangle }\text{,}
\end{equation}%
and
\begin{equation}
\langle \langle \sigma _{k}^{-}|\sigma _{k}^{+}\rangle \rangle _{\omega }=%
\frac{-\langle \sigma ^{z}\rangle (\omega -\Omega _{k})}{(\omega -\Omega
_{k})(\omega -\omega _{A})+g^{2}\langle \sigma ^{z}\rangle }\text{.}
\end{equation}%
There exist two poles $\omega =\omega _{k}^{(+)}$ and $\omega =\omega
_{k}^{(-)}$:
\begin{equation}
\omega _{k}^{(\pm )}=\Omega _{D}\pm \epsilon _{k},
\end{equation}%
which are the dispersion relations. Here
\begin{align}
\Omega _{D}& =\frac{1}{2}(\omega _{A}+\Omega _{k})\text{,}  \label{dis01} \\
\epsilon _{k}& =\frac{1}{2}\sqrt{(\Omega _{k}-\omega _{A})^{2}-4g^{2}\langle
\sigma ^{z}\rangle }\text{.}  \notag
\end{align}

We analyze the properties of the retarded Green functions of $\langle
\langle \hat{a}_{k}|\hat{a}_{k}^{\dag }\rangle \rangle _{\omega
+i\varepsilon }$ and $\langle \langle \sigma _{k}^{-}|\sigma _{k}^{+}\rangle
\rangle _{\omega +i\varepsilon }$ by decomposing them into two branches of
wave, respectively. The propagating photons have the form as
\begin{equation}
\langle \langle \hat{a}_{k}|\hat{a}_{k}^{\dag }\rangle \rangle _{\omega
+i\varepsilon }=A_{k}G^{+}(k,\omega )+B_{k}G^{-}(k,\omega ),
\end{equation}%
where the free Green functions are denoted by
\begin{equation}
G^{\pm }(k,\omega )=\frac{1}{\omega -\omega _{k}^{(\pm )}+i\varepsilon }%
\text{,}
\end{equation}%
and the atomic part has the form as
\begin{equation*}
\langle \langle \sigma _{k}^{-}|\sigma _{k}^{+}\rangle \rangle _{\omega
+i\epsilon }=-\langle \sigma ^{z}\rangle \lbrack B_{k}G^{+}(k,\omega
)+A_{k}G^{-}(k,\omega )]\text{,}
\end{equation*}%
where
\begin{equation}
A_{k}=\frac{\omega _{k}^{(+)}-\omega _{A}}{\omega _{k}^{(+)}-\omega
_{k}^{(-)}},B_{k}=\frac{\omega _{A}-\omega _{k}^{(-)}}{\omega
_{k}^{(+)}-\omega _{k}^{(-)}}\text{,}
\end{equation}%
are the amplitudes of the two wave branches.

Transforming the Green functions above back to the real time representation,
one can observe that the photons propagate with two frequencies $\omega
_{k}^{(\pm )}$. If we regard the transmission of the localized photons as
propagating wave, the two wave branches or two partial waves with $\omega
_{k}^{(\pm )}$ have the amplitudes $A_{k}$ and $B_{k}$, respectively. For $%
\langle \langle \sigma _{k}^{-}|\sigma _{k}^{+}\rangle \rangle _{\omega }$
it has been observed that \cite{sup} the total collection of the identical
two-level atoms can be regarded as an ensemble of $N$ spins and thus its
collective excitation can be described as spin waves, which are
characterized by $\langle \langle \sigma _{k}^{-}|\sigma _{k}^{+}\rangle
\rangle _{\omega }$. It is shown that the spin wave has two eigenfrequencies
$\omega _{k}^{(\pm )}$, but with the twist amplitudes $B_{k}$ and $A_{k}$,
respectively. (The detailed analysis is given in the appendix B.)

For the photons in the CROW from localized modes to propagating modes, we
can now visualize the two wave branches as quasi-particle excitations by
considering the existence of isolated poles of $\langle \langle \hat{a}_{k}|%
\hat{a}_{k}^{\dag }\rangle \rangle _{\omega }$. Suppose that there are such
poles $\omega _{k}^{(\pm )}\rightarrow \omega _{k}^{(\pm )}-i\gamma _{\pm }$
on the complex plane. They correspond to the life $1/\gamma _{\pm }$ of the
quasi-particle excitations characterizing the two branches of propagating
wave in the coupled cavity array. By phenomenologically adding imaginary
parts $-i\gamma _{c}$ and $-i\gamma _{A}$ to the cavity eigenfrequency $%
\omega _{C}$ and the atomic level spacing $\omega _{A}$ respectively, $%
\gamma _{\pm }$ can be explicitly expressed obviously \ in terms of $%
-i\gamma _{c}=$ $-i\kappa $ and $-i\gamma _{A}=-i\gamma $ (the details in
the next section). This means that the decay of transporting photons is just
induced by the cavity decay and the atom natural line width.

\section{Coherent transmission of\ photons with slowed group velocity}

From the dispersion relation (\ref{dis01}), it can be observed that the
population inversion $\langle \sigma ^{z}\rangle $ can directly affect the
basic features of coherent transmission of photons in the CROW. To enhance
this influence, we put more doping atoms in a cavity. Suppose that every
cavity is doped by $n$ identical atoms without interaction among themselves.
Then the parts of the Hamiltonian in section II concerning atoms become
\begin{equation}
H_{A}=\sum_{\alpha=0}^{N-1}\frac{\omega _{A}}{2}S_{\alpha}^{z},H_{AC}=g\sum%
\limits_{\alpha=0}^{N-1}\hat{a}_{\alpha}S_{\alpha}^{+}+h.c.\text{,}  \notag
\end{equation}%
where we have introduced the collective spin $\mathbf{S}_{\alpha}=%
\sum_{l=1}^{n}\mathbf{\sigma }_{\alpha l}$. In this sense the above
frequencies $\omega _{k}^{(\pm )}$ can be modified by replacing $\langle
\sigma ^{z}\rangle $ with
\begin{equation}
\langle S^{z}\rangle =\frac{1}{N}\sum_{\alpha=0}^{N-1}\sum_{l}^{n}\langle
\sigma _{\alpha l}^{z}\rangle \text{.}
\end{equation}%
Obviously, the average of the total spin is bounded as $-n\leq \langle
S^{z}\rangle \leq n$.

Before discussing the group velocity of photons, we investigate the change
of bandwidth of this photonic-crystal-like system. Because the group
velocity $v_{g}^{k}$ can be calculated according to $v_{g}^{k}=d\omega /dk$,
which concerns the range of $\omega$, the change of bandwidth plays an
important role. Without the doped atoms the spectrum of photons should have
only one band, and the central line should be at $\omega _{C}$. However when
atoms are doped, the spectrum splits into two bands with eigenfrequencies $%
\omega _{k}^{(\pm )}$. Then the central lines shift to $\Omega _{D}\pm
\epsilon _{k=\pi /2\ell }$. Without population inversion, i.e.,$\langle
S^{z}\rangle <0$, the two bands have the same width $W=\left\vert \omega
_{k}^{(\pm )}|_{k=0}-\omega _{k}^{(\pm )}|_{k=\pi /2\ell }\right\vert <2J$,
which is calculated as
\begin{equation}
W=|F_{\pm }(\left\langle S_{0}^{z}\right\rangle ,0)-F_{\pm }(\left\langle
S_{0}^{z}\right\rangle ,\frac{\pi }{2\ell })|\text{,}
\end{equation}%
where

\begin{equation}
F_{\pm }(x,k)=\sqrt{\frac{1}{4}\left[ \delta +J\cos (k\ell )\right]
^{2}-g^{2}x}\text{,}
\end{equation}%
for $\delta =\omega _{C}-\omega _{A}$. This means that the bandwidth becomes
narrower when cavities are coupled to more atoms.

Next we consider the group velocity of photon propagation
\begin{equation}
v_{g}^{(\pm )k}=J\ell \sin (k\ell )\left[ 1\pm \frac{\delta +2J\cos (k\ell )%
}{2F_{\pm }(\langle S^{z}\rangle ,k)}\right] \text{,}  \label{ad:1}
\end{equation}%
for various cases. At $k=\pi /2\ell $, the group velocities of $\omega _{\pi
/2\ell }^{(\pm )}$ read
\begin{equation}
v_{g}^{(\pm )\pi /2\ell }=J\ell \left[ 1\pm \frac{\delta }{\varkappa }\right]
\text{,}  \label{ad:2}
\end{equation}%
and the amplitudes of the photon propagator can be calculated as
\begin{equation}
A_{\pi /2\ell }=\frac{\delta +\varkappa }{2\varkappa }\text{, }B_{\pi /2\ell
}=\frac{-\delta +\varkappa }{2\varkappa }\text{,}
\end{equation}%
respectively, where $\varkappa =\sqrt{\delta ^{2}-4g^{2}\langle S^{z}\rangle
}$\emph{.}
\begin{figure}[h]
\begin{center}
\includegraphics[width=8cm,height=8cm]{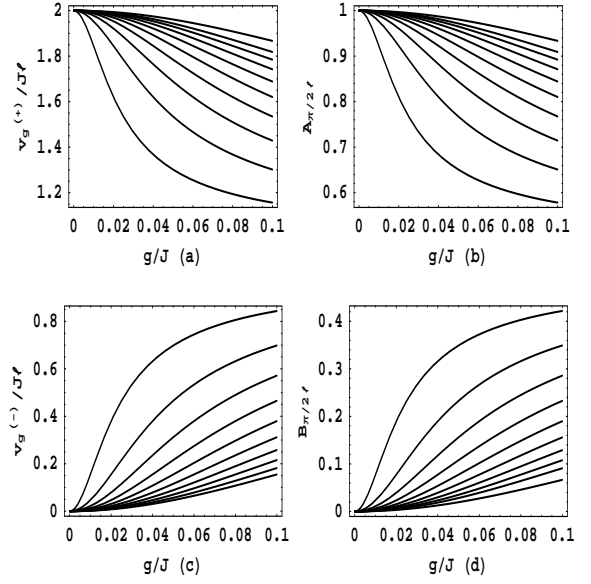}
\end{center}
\caption{There are ten curves with $J\geq \protect\delta \geq 0.1J$ in the
fig(a)-(d), respectively. The group velocities of $v_{g}^{(+)}$, $%
v_{g}^{(-)} $ and the amplitudes $A_{k}$, $B_{k}$ are functions of coupling
strength $g$ at $k=\protect\pi /(2\ell )$. In the fig(a)-(d) the lattice
constant $\ell =1 $, $|<S^{z}>|=10$. In fig(a) and fig(b) the upper curves
are for $\protect\delta =J$, while in fig(c) and fig(d) the upper curves are
for $\protect\delta =0.1J$.}
\end{figure}
%%%%%%%%%%%%%%%%%%%%%%%%%%%%%%%%%%%%%%%%%%%%%%%%%%%%%%%%%%%%%%%%%%%%%%%%%%%%%
We now consider the case with most atoms in the ground state, i.e., $\langle
S^{z}\rangle <0$. \ When $\delta \gg 2g\sqrt{|\langle S^{z}\rangle |}$, the
amplitude at the band center $A_{\pi /2\ell }\rightarrow 1$, $B_{\pi /2\ell
}\rightarrow 0$, and then
\begin{equation}
\langle \langle \hat{a}_{\pi /2\ell }|\hat{a}_{\pi /2\ell }^{\dag }\rangle
\rangle _{\omega +i\varepsilon }\approx \frac{1}{\omega -\omega _{\pi /2\ell
}^{(+)}+i\varepsilon }.
\end{equation}
In other words, in this limit the photon modified by the atoms tends to have
an eigenfrequency $\omega _{\pi /2\ell }^{(+)}$. Correspondingly the group
velocity reaches its maximum $v_{g}^{(+)\pi /(2\ell )}\approx 2J\ell $, and
the quasi-spin wave for the atomic excitations is characterized by the Green
function
\begin{equation}
\langle \langle \sigma _{\pi /2\ell }^{-}|\sigma _{\pi /2\ell }^{+}\rangle
\rangle _{\omega +i\varepsilon }\approx \frac{|\langle S^{z}\rangle |}{%
\omega -\omega _{\pi /2\ell }^{(-)}+i\varepsilon },
\end{equation}%
which has a distinct eigenfrequency $\omega _{\pi /2\ell }^{(-)}$. With $%
g\rightarrow 0$, $\omega _{\pi /2\ell }^{(+)}$ and $\omega _{\pi /2\ell
}^{(-)}$ approach $\omega _{C}$ and $\omega _{A}$, respectively. On the
contrary, when $\delta \ll -2g\sqrt{|\langle S^{z}\rangle |}$, we make a
similar argument: as $A_{\pi /2\ell }\rightarrow 0$, $B_{\pi /2\ell
}\rightarrow 1$, the photons and atomic spin wave propagate with
eigenfrequencies $\ \omega _{\pi /2\ell }^{(-)}$ and $\omega _{\pi /2\ell
}^{(+)}$, respectively. By letting $g\rightarrow 0$, $\omega _{\pi /2\ell
}^{(+)}$ and $\omega _{\pi /2\ell }^{(-)}$ can be revived as $\omega _{A}$
and $\omega _{C}$, respectively. The group velocity of photons can also
reach its maximum $v_{g}^{(-)\pi /2\ell }\approx 2J\ell $ (The detailed
analysis is given in the appendix B). These observations are different from
the results obtained in the simple cavity - cavity coupling system without
atom doping in Ref.\cite{Fanprl04} .

\begin{figure}[h]
\begin{center}
\includegraphics[width=9cm,height=4cm]{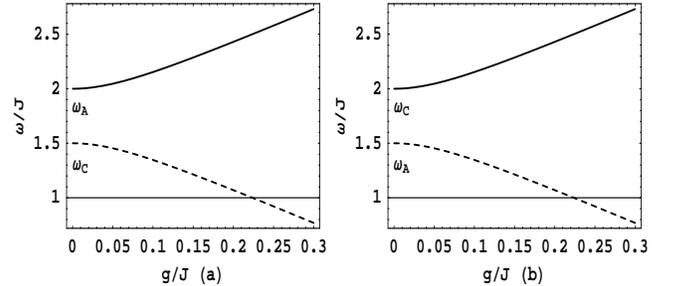}
\end{center}
\caption{In fig (a) and (b), $\protect\omega _{\protect\pi /(2\ell )}^{(+)}$%
(solid) and $\protect\omega _{\protect\pi /(2\ell )}^{(-)}$(dashed) are
functions of coupling strength g and $|<S^{z}>|=10$. In fig (a) $\protect%
\omega _{C}=2J$, $\protect\omega _{A}=1.5J$ and $\protect\delta =0.5J$,
while in fig (b) $\protect\omega _{C}=1.5J$, $\protect\omega _{A}=2J$ and $%
\protect\delta =-0.5J$.}
\label{fig4}
\end{figure}

Analyzing the features of eigenfrequencies $\omega _{k}^{(\pm )}$ for photon
and atom parts in the case of weak coupling, we have observed that $\omega
_{k}^{(\pm )}$ have different preferences to approach frequencies of pure
photons or bare atoms. It is concluded from this observation that , if $%
\delta >0$, photons prefer $\omega _{k}^{(+)}$ \ while the atomic spin wave
prefers $\omega _{k}^{(-)}$ ; if $\delta <0$,\ the conclusion is just on the
contrary. We illustrate these analysis in Fig. \ref{fig4}.
%%%%%%%%%%%%%%%%%%%%%%%%%%%%%%%%%%%%%%%%%%%%%%%%%%%%%%%%%%%%%%%%%%%%%%%%%%%%%

Next we consider how a coherent pump induces population inversion to result
in a laser-like output for the CROW.\emph{\ }In the discussions above, we
have considered an ideal case in which the quantum dissipation and dephasing
due to the influence of the environment are neglected. Meanwhile, we have
not concerned the role of $\langle S^{z}\rangle $, the average value of
total atoms population. Here, we also assumed that $\langle S^{z}\rangle $
is not tunable . But for an open system, $\langle S^{z}\rangle $ becomes a
time-dependent parameter. We can tune the population of atoms to change the
properties of photon transmission. When the population inversion takes
place, we expect that laser-like output emerges.

To see more details, let us consider a realistic case that the cavity damp
has the same rate $\kappa $ and the atoms have the decay rate $\gamma $ due
to spontaneous radiation. Then $\omega _{C}\rightarrow \omega _{C}-i\kappa $
and $\omega _{A}\rightarrow \omega _{A}-i\gamma $, \ so the eigenfrequencies
of photonic band become
\begin{equation}
\omega _{k}^{(\pm )}=\Omega _{D}-\frac{i\left( \gamma +\kappa \right) }{2}%
\pm \Omega _{\pm }\text{,}
\end{equation}%
where we define $\Lambda =\gamma -\kappa $ and
\begin{equation}
\Omega _{\pm }=\frac{1}{2}\sqrt{[\left( \Omega _{k}-\omega _{A}\right)
-i\Lambda ]^{2}-4g^{2}\langle S^{z}\rangle }\text{.}
\end{equation}

For the sake of simplicity, we consider a special case that is $\Omega
_{k}\approx \omega _{A}$, and $\gamma \gg \kappa $. Then the
eigenfrequencies read
\begin{equation}
\omega _{k}^{(\pm )}=\Omega _{D}\pm \frac{i}{2}[f(\langle S^{z}\rangle
)\gamma \mp \gamma \mp \kappa ]\text{,}
\end{equation}%
where $f(x)=\sqrt{1+4xg^{2}/\gamma ^{2}}$. If the imaginary part $\lambda $
of an eigenfrequency (e.g., $\omega _{k}^{(+)}$) is positive, the laser-like
output will appear since a component $A_{k}/(\omega -\omega _{k}^{(+)})$ of $%
\langle \langle \hat{a}_{k}|\hat{a}_{k}^{\dag }\rangle \rangle _{\omega }$
has a real time correspondence

\begin{equation}
A(t)=\int \frac{A_{k}d\omega }{\omega -\omega _{k}^{(+)}}\sim -i\theta
(t)e^{-i\Omega _{D}t+\lambda t}\text{,}
\end{equation}%
where $A_{k}$ is a slowly varying amplitude in the $k$-space. Actually, when
most of atoms stay at the ground state, i.e. $\langle S^{z}\rangle \leq 0$,
it is impossible for $\omega _{k}^{(+)}$ and $\omega _{k}^{(-)}$ to have a
positive imaginary part; when the population of most atoms inverts, i.e. $%
\langle S^{z}\rangle >0$, the imaginary part of $\omega _{k}^{(+)}$ may be
positive.

Furthermore there is a threshold value of $\langle S^{z}\rangle $ to satisfy
the condition $[f(\langle S^{z}\rangle )-1]\gamma >\kappa $. When $\langle
S^{z}\rangle \ll \gamma ^{2}/(4g^{2})$, we explicitly obtain the threshold
value of $\left\langle S^{z}\right\rangle _{T}$ as
\begin{equation}
\langle S^{z}\rangle _{T}=\frac{\gamma \kappa }{2g^{2}}\text{.}
\end{equation}%
Above this threshold value, the eigenfrequency
\begin{equation}
\omega _{k}^{(+)}\sim \Omega _{D}+i(\frac{g^{2}}{\gamma }\langle
S^{z}\rangle -\frac{\kappa }{2})
\end{equation}%
has a positive imaginary part, which results in a laser-like output in the
CROW. It is very interesting that the $\langle S^{z}\rangle _{T}$ is very
similar to the threshold value of population inversion in the generic laser
theory. We also notice that, in weak-coupling limit, $\omega _{k}^{(-)}$ can
not be a robust frequency of laser-like output since it damps fast with rate
$\left[ f\left( \langle S^{z}\rangle \right) +1\right] \gamma +\kappa $.

\section{Susceptibility analysis for light propagation in the doped CROW}

The above analysis displays a possibility to implement a slow light
propagation in the doped CROW, but the calculation of the group velocity
from the dispersion relation shows that, only for certain wave vector $k,$
can the group velocity be reduced down. Thus for the propagation of a wave
packet or a light pulse, we still need some details for the absorption and
dispersion of light in the doped CROW. We use the dynamic algebraic method
developed for the atomic ensemble based on quantum memory with EIT \cite%
{sunprl91}. The original method was proposed for the conventional EIT
system, which consists of a vapor cell with three-level $\Lambda -$atoms
near resonantly coupled to the controlling and quantized probe light. Our
dynamic symmetry analysis is based on the hidden dynamic symmetry described
by the semi-direct product of quasi-spin $SU(2)$ and the boson algebra of
the excitations. This method allows us to build a dynamic equation
describing the propagation of the probe light in this atomic ensemble with
atomic collective excitations \cite{liyong}.

Now we apply this algebraic method to calculate the susceptibility of light
for the group velocity of photonic wave packet propagating along the doped
CROW. Then we investigate how the susceptibility depends on the various
control parameters.

We simplify our model by using the collective operators $b_{\alpha}=S_{%
\alpha}^{-}/\sqrt{n}$ and $b_{\alpha}^{\dag }=S_{\alpha}^{+}/\sqrt{n}$,
which represent the quasi-spin wave in the low-excitation limit that only
few atoms populated in their excited state. In this case we can check that
the spin wave is bosonic excitation since the boson commutation relation $%
[b_{\alpha},b_{\alpha^{\prime }}^{\dag }]=\delta _{\alpha,\alpha^{\prime }}$
is satisfied in the low-excitation limit. Then the total Hamiltonian for the
hybrid system with many-atom doping becomes
\begin{align}
H& =\omega _{C}\sum_{\alpha}a_{\alpha}^{\dag
}a_{\alpha}+J\sum_{\alpha}\left( a_{\alpha}^{\dag
}a_{\alpha+1}+a_{\alpha+1}^{\dag }a_{\alpha}\right)  \notag \\
& +\omega _{A}\sum_{\alpha}b_{\alpha}^{\dag }b_{\alpha}+\sum_{\alpha}g\sqrt{n%
}\left( a_{\alpha}^{\dag }b_{\alpha}+b_{\alpha}^{\dag }a_{\alpha}\right) .
\end{align}%
Its $k$-space representation $H=\sum_{k}$ $H_{k}$ is a simple sum of the $k$%
-component
\begin{equation}
H_{k}=\Omega _{k}a_{k}^{\dag }a_{k}+\omega _{A}b_{k}^{\dag }b_{k}+g\sqrt{n}%
\left( a_{k}^{\dag }b_{k}+h.c\right) .
\end{equation}%
Here, we have used the Fourier transformation $b_{k}=\sum_{\alpha}\exp
(ik\ell \alpha)b_{\alpha}/\sqrt{N}$.

For each mode $k$,\ we can write down the Heisenberg equations of operators $%
a_{k}$ and $b_{k}$.
\begin{align*}
i\partial _{t}a_{k}& =-i\kappa a_{k}+\Omega _{k}a_{k}+g\sqrt{n}b_{k}, \\
i\partial _{t}b_{k}& =-i\gamma b_{k}+\omega _{A}b_{k}+g\sqrt{n}a_{k}.
\end{align*}%
Here, we have phenomenologically introduced the decay rates $\kappa $ and $%
\gamma $, and $\gamma \gg \kappa $. In the interaction picture, we adopt the
time-dependent transformation,
\begin{equation}
a_{k}=\widetilde{a}_{k}e^{-i\Omega _{k}t},b_{k}=\widetilde{b}_{k}e^{-i\Omega
_{k}t}
\end{equation}%
for $a_{k}$ and $b_{k}$ to remove the fast varying parts of the light field
and the atomic collective excitations. Then the above equations of motion
are reduced into
\begin{align*}
\partial _{t}\widetilde{a}_{k}& =-\kappa \widetilde{a}_{k}-ig\sqrt{n}%
\widetilde{b}_{k}, \\
\partial _{t}\widetilde{b}_{k}& =-\gamma \widetilde{b}_{k}-i(\omega
_{A}-\Omega _{k})b_{k}-ig\sqrt{n}\widetilde{a}_{k}.
\end{align*}

In general the steady state solution of the equations above determines the
susceptibility of photon transmission. It is noticed that the quantized
light described by $a_{k}$ is the superposition of some localized modes $%
a_{j}$. On the contrary, the spatially distributed photon field is
characterized by $a_{\alpha}=\sum_{k}\exp (-ik\ell \alpha)a_{k}/\sqrt{N},$%
which means the inhomogeneous polarization $\left\langle
P_{\alpha}\right\rangle $ depends on the spatial position. Correspondingly,
we have the $k$-space representation of the light field
\begin{equation}
E_{k}(t)=\sqrt{\frac{\omega _{C}}{2V\epsilon _{0}}}a_{k}e^{-i\Omega
_{k}t}+h.c.
\end{equation}%
In comparison with the classical expression $E_{k}(t)=\epsilon _{k}\exp
(-i\Omega _{k}t)+h.c.,$it is recognized that
\begin{equation*}
\epsilon _{k}\sim \sqrt{\frac{\omega _{C}}{2V\epsilon _{0}}}a_{k}.
\end{equation*}%
On the other hand the linear response of medium is described by the local
polarization $\left\langle P_{k}\right\rangle =\left\langle
p_{k}\right\rangle \exp (-i\Omega _{k}t)+h.c.$, where the average
polarization
\begin{equation}
\left\langle p_{k}\right\rangle =\frac{\mu }{V}\sqrt{n}\langle \widetilde{b}%
_{k}\rangle
\end{equation}%
slowly varies \ and determined by an average value of excitation operator $%
\widetilde{b}_{k}$; $\mu $ denotes the dipole moment of single atom, and $V$
is the effective mode volume\cite{bqoptics}. It is related to the
susceptibility $\chi _{k}$ of the $k$-space by
\begin{equation}
\chi _{k}=\frac{\left\langle p_{k}\right\rangle }{\left\langle \epsilon
_{k}\right\rangle \epsilon _{0}}=\frac{\sqrt{n}\mu }{\left\langle \epsilon
_{k}\right\rangle \epsilon _{0}V}\langle \widetilde{b}_{k}\rangle
\end{equation}
since $\left\langle P_{k}\right\rangle =\chi _{k}E_{k}(t).$

To calculate the susceptibility in our case, we need the steady state
solution satisfying $\partial _{t}\widetilde{b}_{k}=0$, or
\begin{equation*}
\gamma \widetilde{b}_{k}+i(-\delta -2J\cos k\ell)\widetilde{b}_{k}+ig\sqrt{n}%
\widetilde{a}_{k}=0,
\end{equation*}%
for $\gamma \gg \kappa $. Here, $\delta =\omega _{C}-\omega _{A}$ is the
detuning between photons and atoms. In the steady state approach, we can
take the expectation value for the above equation
\begin{equation*}
ig\sqrt{n}\left\langle \widetilde{a}_{k}\right\rangle =-(\gamma -i\delta
-2J\cos k\ell)\langle \widetilde{b}_{k}\rangle .
\end{equation*}%
Since the dipole approximation $g=-\mu \sqrt{\omega _{C}/(2V\epsilon _{0})}$%
, the linear susceptibility
\begin{align*}
\chi _{k}& \equiv \chi _{1k}+i\chi _{2k} \\
& =\frac{2ig^{2}n}{\omega _{C}[\gamma -i(\delta +2J\cos k\ell)]}\text{.}
\end{align*}%
The real part
\begin{equation}
\chi _{1k}=\frac{-(\delta +2J\cos k\ell)2g^{2}n}{\omega _{C}[\gamma
^{2}+(\delta +2J\cos k\ell)^{2}]}\text{,}
\end{equation}%
and the imaginary part
\begin{equation}
\chi _{2k}=\frac{2g^{2}n\gamma }{\omega _{C}[\gamma ^{2}+(\delta +2J\cos
k\ell)^{2}]}  \label{absorption}
\end{equation}%
are related to the dispersion and absorption of the light field in the CROW,
respectively.

\begin{figure}[h]
\begin{center}
\includegraphics[width=8cm,height=8cm]{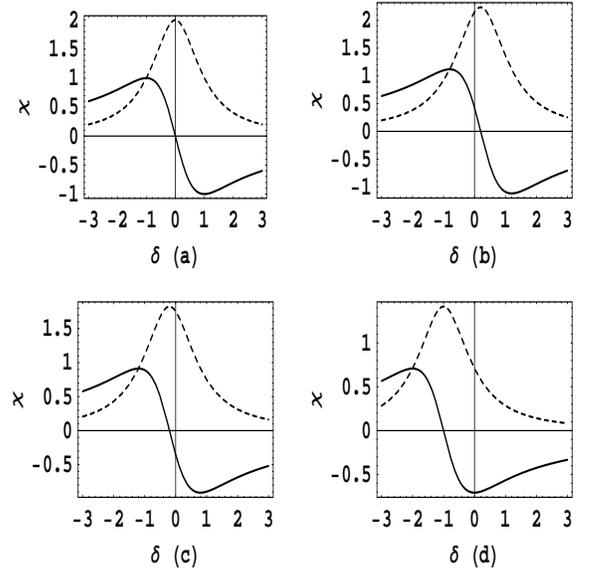}
\end{center}
\caption{Real part $\protect\chi_{1}$ (solid) and imaginary part $\protect%
\chi_{2}$ (dashed) of the susceptibility $\protect\chi$ vs the light
detuning $\protect\delta$ in normalized units of $\protect\gamma$ according
to: (a) $k=\protect\pi /2\ell $ and $J=0.1$; (b) $k=\protect\pi /\ell $ and $%
J=0.1 $; (c) $k=0$ and $J=0.1$; (d) $k=0$ and $J=0.8$. The other parameters
are given as: $\protect\omega_{C}=1$ $g\protect\sqrt{n}=1$.}
\label{fig5}
\end{figure}

Because the photons have the band structure, the properties of dispersion
and absorption of photons vary with the the wave vector with momentum index $%
k$. While the real part $\chi _{1k}$ reaches its maximum at $\delta
=-(\gamma +2J\cos k\ell )$, the imaginary part $\chi _{2k}$ reaches its
maximum at $\delta =-2J\cos k\ell $ and thus the absorption is a
considerable property of the system. Here, the photons with different wave
vectors $k$ will have different character of absorption. Figure \ref{fig5}
(a)-(c) show the dependence of $\chi _{1k}$ and $\chi _{2k}$ on $k$. The
maximums of the absorption for $k=\pi /2\ell $, $\pi /\ell $, $0$, appear at
three different values of $\delta $. The reason for this phenomena is that
the inter-cavity interaction with the coupling constant $J$ will shift the
resonance point in general, but if $k=\pi /2\ell $, the coupling has no
effect on for spectral structure. In view of our analysis, the absorption
directly depends on the wave vector $k$. At the same time we can imagine
that the character of absorption can influence the group velocity of
photons. It is obvious to see that an unavoidable loss effect appears for
the group velocity when the atom media absorbs light strongly. The
dispersion relation of photons is described by   $\Omega _{k}=\omega
_{C}+2J\cos {k\ell }$ , from which we calculate the group velocity as%
\begin{equation}
v_{g}^{k}=|\frac{d\Omega _{k}}{dk}|\propto 2J\ell .
\end{equation}
But from  Eq.(\ref{absorption}) it can be seen that the media
absorption characterized by $\chi _{2k}$, will be stronger when
$2J\ell $ becomes smaller. In other words,  the considerable
absorption corresponds to a slow group velocity. Actually, since the
effect of the group velocity  is due to  a spectrum structure of the
wave vector $k$, only  a small range of $k$  around the point
corresponding to the minimum group velocity, avoided is the  higher
order dispersion. The the point of minimum group velocity and the
point of maximum absorption are related and somewhat close. This
fact means that there is some unavoidable loss.

\section{Conclusion}

We have studied the coherent transmission of photons with local modes along
the CROW coupled to artificial two-level atoms. Under the weak-coupling
limit, we use the stimulated Raman excitation to tune the level spacing of
the effective two-level system so that the properties of photons in the CROW
can be manipulated coherently. As the above results display, if we prepare
the hybrid system as that $\omega _{C}\gg \omega _{A}$ or $\omega _{C}\ll
\omega _{A},$ the group velocity of photons in the doped CROW will reach
maximum under the two cases. Meanwhile the two eigenfrequencies of the
hybrid system have preference that, while one tends to the frequency of
photons, another tends to that of quasi-spin wave of the total atoms. By
controlling the average population of the doping atoms in the CROW with
decay, we predict that the laser-like output may occur. With such an exotic
photonic band structure, the light with different $k$ has different
properties of absorption.

In \cite{zhou} about control of photon transmission in CROW by doping
artificial atoms for various hybrid structures, we study the case of the
resonate three-level doping atoms by making use of the quasi-spin wave
theory based on a mean field method. This investigation will make a
corporate effort for the coherent transmission of photons in an artificial
structure, where both the EIT effect and the band-like structure are
utilized simultaneously.

This work is supported by the NSFC with grant No.90403018, 10375038,
60433050, and NKBRPC with No.2004CB318000, and NFRPC with No. 2001CB309310
and 2005CB724508. One (LZ) of authors gratefully acknowledges the support of
K. C. Wong Education Foundation, Hong Kong. \appendix

\section{ Equations of motion for the two time Green function}

\label{app:appendix a} \bigskip In this appendix, we provide a detailed
derivation of the approximately closed system of the equations for two-time
Green functions as following text \ \ \
\begin{align}
& \langle \langle \hat{a}_{k}|\hat{a}_{k}^{\dag }\rangle \rangle _{\omega }%
\text{, }\langle \langle \sigma _{k^{\prime }}^{-}|\sigma _{k}^{+}\rangle
\rangle _{\omega }\text{, }\langle \langle \sigma _{k}^{-}|\sigma
_{k}^{+}\rangle \rangle _{\omega },  \notag \\
& \langle \langle \hat{a}_{k}|\sigma _{k^{\prime }}^{+}\rangle \rangle
_{\omega }\text{, and }\langle \langle \sigma _{k}^{-}|\hat{a}_{k^{\prime
}}^{\dag }\rangle \rangle _{\omega }.
\end{align}%
From the commutation relation between $\hat{a}_{k}$ and $H$, the equation of
$\langle \langle \hat{a}_{k}|\hat{a}_{k}^{\dag }\rangle \rangle _{\omega }$
is obtained as
\begin{equation*}
\omega \langle \langle \hat{a}_{k}|\hat{a}_{k}^{\dag }\rangle \rangle
_{\omega }=\langle \lbrack \hat{a}_{k},\hat{a}_{k}^{\dag }]\rangle +\langle
\langle \lbrack \hat{a}_{k},H]|\hat{a}_{k}^{\dag }\rangle \rangle _{\omega },
\end{equation*}%
or
\begin{equation}
(\omega -\Omega _{k})\langle \langle \hat{a}_{k}|\hat{a}_{k}^{\dag }\rangle
\rangle _{\omega }=1+g\langle \langle \sigma _{k}^{-}|\hat{a}_{k}^{\dag
}\rangle \rangle _{\omega }.
\end{equation}%
Since the above equation concerns the two-time Green function $\ \langle
\langle \sigma _{k}^{-}|\hat{a}_{k}^{\dag }\rangle \rangle _{\omega },$ we
need its motion equation, but here, we first calculate the equation of$%
\langle \langle \sigma _{k^{\prime }}^{-}|\sigma _{k}^{+}\rangle \rangle
_{\omega }$
\begin{equation*}
\omega \langle \langle \sigma _{k^{\prime }}^{-}|\sigma _{k}^{+}\rangle
\rangle _{\omega }=\langle \lbrack \sigma _{k^{\prime }}^{-},\sigma
_{k}^{+}]\rangle +\langle \langle \lbrack \sigma _{k^{\prime
}}^{-},H]|\sigma _{k}^{+}\rangle \rangle _{\omega }\text{,}
\end{equation*}%
or
\begin{align}
\omega \langle \langle \sigma _{k^{\prime }}^{-}|\sigma _{k}^{+}\rangle
\rangle _{\omega }& =\omega _{A}\langle \langle \sigma _{k^{\prime
}}^{-}|\sigma _{k}^{+}\rangle \rangle _{\omega } \\
& -\frac{g}{N}\sum_{k^{\prime \prime }\alpha}e^{-i(k^{\prime }-k^{\prime
\prime })\ell\alpha}\langle \langle \sigma _{\alpha}^{z}\hat{a}_{k^{\prime
\prime }}|\sigma _{k}^{+}\rangle \rangle _{\omega }  \notag \\
& -\frac{1}{N}\sum_{\alpha=0}^{N-1}e^{-i(k^{\prime }-k)\ell\alpha}\langle
\sigma _{\alpha}^{z}\rangle .  \notag
\end{align}%
\emph{The mean\ field approximation} \ assumes that $\langle \sigma
_{\alpha}^{z}\rangle $\ can be factorized from $\langle \langle \sigma
_{\alpha}^{z}\hat{a}_{k^{\prime \prime }}|\sigma _{k^{\prime }}^{+}\rangle
\rangle _{\omega }$, i.e.,%
\begin{equation}
\langle \langle \sigma _{\alpha}^{z}\hat{a}_{k^{\prime \prime }}|\sigma
_{k^{\prime }}^{+}\rangle \rangle _{\omega }\approx \langle \sigma
_{\alpha}^{z}\rangle \langle \langle \hat{a}_{k^{\prime \prime }}|\sigma
_{k^{\prime }}^{+}\rangle \rangle _{\omega }
\end{equation}%
and then the above Green function hierarchy is cut off. Thus we get a system
of Green function equations
\begin{align*}
\omega \langle \langle \sigma _{k^{\prime }}^{-}|\sigma _{k}^{+}\rangle
\rangle _{\omega }& \approx -\frac{g}{N}\sum_{k^{\prime \prime
}\alpha}e^{-i(k^{\prime }-k^{\prime \prime })\ell\alpha}\langle \sigma
_{\alpha}^{z}\rangle \langle \langle \hat{a}_{k^{\prime \prime }}|\sigma
_{k}^{+}\rangle \rangle _{\omega } \\
& \frac{-1}{N}\sum_{\alpha=0}^{N-1}e^{-i(k^{\prime }-k)\ell\alpha}\langle
\sigma _{\alpha}^{z}\rangle +\omega _{A}\langle \langle \sigma _{k^{\prime
}}^{-}|\sigma _{k}^{+}\rangle \rangle _{\omega },
\end{align*}%
or
\begin{align}
\langle \langle \sigma _{k^{\prime }}^{-}|\sigma _{k}^{+}\rangle \rangle
_{\omega }& =-\frac{1}{N\left( \omega -\omega _{A}\right) }%
[\sum_{\alpha=0}^{N-1}e^{-i(k^{\prime }-k)\ell\alpha}\langle \sigma
_{\alpha}^{z}\rangle \\
& +g\sum_{k^{\prime \prime }\alpha}e^{-i(k^{\prime }-k^{\prime \prime
})\ell\alpha}\langle \sigma _{\alpha}^{z}\rangle \langle \langle \hat{a}%
_{k^{\prime \prime }}|\sigma _{k}^{+}\rangle \rangle _{\omega }].  \notag
\end{align}%
For $k=k^{\prime }$, \ the \ Green function $\left\langle \left\langle
\sigma _{k}^{-}|\sigma _{k}^{+}\right\rangle \right\rangle _{\omega }$ \
satisfies
\begin{align}
\left\langle \left\langle \sigma _{k}^{-}|\sigma _{k}^{+}\right\rangle
\right\rangle _{\omega }& =-\frac{1}{N\left( \omega -\omega _{A}\right) }%
[\sum_{\alpha=0}^{N-1}\langle \sigma _{\alpha}^{z}\rangle +  \notag \\
& g\sum_{k^{\prime }\alpha}e^{-i(k-k^{\prime })\ell\alpha}\langle \sigma
_{\alpha}^{z}\rangle \langle \langle \hat{a}_{k^{\prime }}|\sigma
_{k}^{+}\rangle \rangle _{\omega }].
\end{align}%
We notice that the equation of $\langle \langle \hat{a}_{k}|\sigma
_{k^{\prime }}^{+}\rangle \rangle _{\omega }$ is given by
\begin{equation*}
\omega \langle \langle \hat{a}_{k}|\sigma _{k^{\prime }}^{+}\rangle \rangle
_{\omega }=\langle \langle \lbrack \hat{a}_{k},H]|\sigma _{k^{\prime
}}^{+}\rangle \rangle _{\omega },
\end{equation*}%
or
\begin{equation}
(\omega -\Omega _{k})\langle \langle \hat{a}_{k}|\sigma _{k^{\prime
}}^{+}\rangle \rangle _{\omega }=g\langle \langle \sigma _{k}^{-}|\sigma
_{k^{\prime }}^{+}\rangle \rangle _{\omega }.
\end{equation}

In order to derive the equation about $\langle \langle \sigma _{k}^{-}|\hat{a%
}_{k^{\prime }}^{\dag }\rangle \rangle _{\omega }$ , we use the second kind
of motion equation (\ref{AB2}),
\begin{equation*}
\omega \langle \langle \sigma _{k}^{-}|\hat{a}_{k^{\prime }}^{\dag }\rangle
\rangle _{\omega }=-\langle \langle \sigma _{k}^{-}|[\hat{a}_{k^{\prime
}}^{\dag },H]\rangle \rangle _{\omega },
\end{equation*}%
or
\begin{equation}
(\omega -\Omega _{k^{\prime }})\langle \langle \sigma _{k}^{-}|\hat{a}%
_{k^{\prime }}^{\dag }\rangle \rangle _{\omega }=g\langle \langle \sigma
_{k}^{-}|\sigma _{k^{\prime }}^{+}\rangle \rangle _{\omega }.
\end{equation}%
By defining
\begin{align*}
\langle \sigma _{k^{\prime }k}^{z}\rangle & =\frac{1}{N}\sum_{%
\alpha=0}^{N-1}e^{-i(k-k^{\prime })\ell\alpha}\langle \sigma
_{\alpha}^{z}\rangle , \\
f_{k}\left( \omega \right) & =(\omega -\omega _{A}+\frac{g^{2}\langle \sigma
^{z}\rangle }{\omega -\Omega _{k}}),
\end{align*}%
the equations about $\langle \langle \sigma _{k}^{-}|\sigma _{k}^{+}\rangle
\rangle _{\omega }$ and $\langle \langle \sigma _{k^{\prime }}^{-}|\sigma
_{k}^{+}\rangle \rangle _{\omega }$ can be finally obtained as
\begin{equation}
f_{k}\left( \omega \right) \langle \langle \sigma _{k}^{-}|\sigma
_{k}^{+}\rangle \rangle _{\omega }=-\langle \sigma ^{z}\rangle
-g^{2}\sum_{k^{\prime }\neq k}\langle \sigma _{k^{\prime }k}^{z}\rangle
\frac{\langle \langle \sigma _{k^{\prime }}^{-}|\sigma _{k}^{+}\rangle
\rangle _{\omega }}{\omega -\Omega _{k^{\prime }}},
\end{equation}

\begin{align}
f_{k^{\prime }}\left( \omega \right) \langle \langle \sigma _{k^{\prime
}}^{-}|\sigma _{k}^{+}\rangle \rangle _{\omega }& =-\langle \sigma
_{k^{\prime }k}^{z}\rangle -\frac{g^{2}\langle \sigma _{kk^{\prime
}}^{z}\rangle \langle \langle \sigma _{k}^{-}|\sigma _{k}^{+}\rangle \rangle
_{\omega }}{(\omega -\Omega _{k})} \\
& -g^{2}\sum_{k^{\prime \prime }\neq k,\neq k^{\prime }}\langle \sigma
_{k^{\prime \prime }k^{\prime }}^{z}\rangle \frac{\langle \langle \sigma
_{k^{\prime \prime }}^{-}|\sigma _{k}^{+}\rangle \rangle _{\omega }}{\omega
-\Omega _{k^{\prime \prime }}}.  \notag
\end{align}

\section{Quasi-spin waves coupled to transferred photons}

\label{app:appendix b}

In this appendix we analyze the physical meaning represented by the Green
functions for photons and atoms that we obtained in section III. First we
explicitly rewrite the coefficients $A_{k}$ and $B_{k}$ in $\langle \langle
\hat{a}_{k}|\hat{a}_{k}^{\dag }\rangle \rangle _{\omega }$ and $\langle
\langle \sigma _{k}^{-}|\sigma _{k}^{+}\rangle \rangle _{\omega }$ as
\begin{align}
A_{k}& =\frac{\omega _{k}^{(+)}-\omega _{A}}{\omega _{k}^{(+)}-\omega
_{k}^{(-)}} \\
& =\frac{\Delta _{k}+\sqrt{\Delta _{k}^{2}-4g^{2}\langle \sigma ^{z}\rangle }%
}{2\sqrt{\Delta _{k}^{2}-4g^{2}\langle \sigma ^{z}\rangle }}\text{,}  \notag
\end{align}%
and
\begin{align}
B_{k}& =\frac{\omega _{A}-\omega _{k}^{(-)}}{\omega _{k}^{(+)}-\omega
_{k}^{(-)}} \\
& =\frac{-\Delta _{k}+\sqrt{\Delta _{k}^{2}-4g^{2}\langle \sigma ^{z}\rangle
}}{2\sqrt{\Delta _{k}^{2}-4g^{2}\langle \sigma ^{z}\rangle }}\text{,}  \notag
\end{align}%
where $\Delta _{k}=\Omega _{k}-\omega _{A}.$ Let $\omega _{C}-\omega
_{A}=\delta .$ When $k=\pi /(2\ell ),$ we obtain $G_{P}\equiv \langle
\langle \hat{a}_{\pi /2\ell }|\hat{a}_{\pi /2\ell }^{\dag }\rangle \rangle
_{\omega }$ as
\begin{align}
G_{P}& =\frac{\delta +\sqrt{\delta ^{2}-4g^{2}\langle \sigma ^{z}\rangle }}{2%
\sqrt{\delta ^{2}-4g^{2}\langle \sigma ^{z}\rangle }}\frac{1}{\omega -\omega
_{\pi /2\ell }^{(+)}}  \notag \\
& +\frac{-\delta +\sqrt{\delta ^{2}-4g^{2}\langle \sigma ^{z}\rangle }}{2%
\sqrt{\delta ^{2}-4g^{2}\langle \sigma ^{z}\rangle }}\frac{1}{\omega -\omega
_{\pi /2\ell }^{(-)}}.
\end{align}

From the above equation we can see that, when $\langle \sigma ^{z}\rangle <0$%
, and $\delta \gg 2g\sqrt{|\langle \sigma ^{z}\rangle |}$, the amplitudes at
the band center
\begin{equation*}
A_{\pi /2\ell }\longrightarrow 1\text{, }B_{\pi /2\ell }\rightarrow 0,
\end{equation*}%
which means
\begin{equation}
\langle \langle \hat{a}_{\pi /2\ell }|\hat{a}_{\pi /2\ell }^{\dag }\rangle
\rangle _{\omega }\approx \frac{1}{\omega -\omega _{\pi /2\ell }^{(+)}}\text{%
,}
\end{equation}%
and the group velocity $v_{g}^{(\pi /2\ell )}\approx 2J\ell $. Meanwhile, if
$g\longrightarrow 0$, the eigenfrequencies of photons and atoms
correspondingly approximate to their original eigenfrequencies without
coupling
\begin{equation}
\omega _{\pi /2\ell }^{(+)}\rightarrow \omega _{C}\text{, }\omega _{\pi
/2\ell }^{(-)}\rightarrow \omega _{A}.
\end{equation}%
If detuning $\delta \ll -2g\sqrt{|\langle \sigma ^{z}\rangle |}$, the values
of amplitudes are in reverse%
\begin{equation}
A_{\pi /2\ell }\longrightarrow 0\text{, }B_{\pi /2\ell }\rightarrow 1\text{.}
\end{equation}%
Thus the Green function of photon at the band center only has one wave
\begin{equation}
\langle \langle \hat{a}_{\pi /2\ell }|\hat{a}_{\pi /2\ell }^{\dag }\rangle
\rangle _{\omega }\approx \frac{1}{\omega -\omega _{\pi /2\ell }^{(-)}}\text{%
.}
\end{equation}%
It also can be obtained that $v_{g}^{(\pi /2\ell )}\approx 2J\ell $.
Meanwhile, we can conclude that, when $g\longrightarrow 0$, the
eigenfrequencies of photon and atom are recovered correspondingly by another
way that
\begin{equation}
\omega _{\pi /2\ell }^{(-)}\rightarrow \omega _{C}\text{, }\omega _{\pi
/2\ell }^{(+)}\rightarrow \omega _{A}.
\end{equation}

Next we study the Green function of the doping atoms $G_{A}\equiv \langle
\langle \sigma _{k}^{-}|\sigma _{k}^{+}\rangle \rangle _{\omega }$:
\begin{equation}
G_{A}=-\langle \sigma ^{z}\rangle \left[ \frac{A_{k}^{\prime }}{\omega
-\omega _{k}^{(+)}}+\frac{B_{k}^{\prime }}{\omega -\omega _{k}^{(-)}}\right]
\end{equation}%
with amplitudes
\begin{equation}
A_{k}^{\prime }=\frac{\omega _{k}^{(+)}-\Omega _{k}}{\omega
_{k}^{(+)}-\omega _{k}^{(-)}}=B_{k}\text{,}
\end{equation}%
and
\begin{equation}
B_{k}^{\prime }=\frac{\Omega _{k}-\omega _{k}^{(-)}}{\omega
_{k}^{(+)}-\omega _{k}^{(-)}}=A_{k}\text{,}
\end{equation}%
which has the similar expression as those of photons. Thus we rewrite the
atomic Green function as
\begin{equation}
G_{A}=-\langle \sigma ^{z}\rangle \lbrack \frac{B_{k}}{\omega -\omega
_{k}^{(+)}}+\frac{A_{k}}{\omega -\omega _{k}^{(-)}}]\text{.}
\end{equation}

We also consider the situation at $k=\pi /(2\ell )$. First we assume $%
\langle \sigma ^{z}\rangle <0$ and $\delta \gg 2g\sqrt{|\langle \sigma
^{z}\rangle |}$, in this case, the value of the amplitudes approximate to
one and zero respectively
\begin{equation}
A_{\pi /2\ell }\longrightarrow 1\text{, }B_{\pi /2\ell }\rightarrow 0\text{,}
\end{equation}%
and thus the Green function of the doping atoms becomes
\begin{equation}
\langle \langle \sigma _{\pi /2\ell }^{-}|\sigma _{\pi /2\ell }^{+}\rangle
\rangle _{\omega }\approx |\langle \sigma ^{z}\rangle |\frac{1}{\omega
-\omega _{\pi /2\ell }^{(-)}}\text{.}
\end{equation}%
However, when $g\longrightarrow 0$,%
\begin{equation}
\omega _{\pi /2\ell }^{(-)}\rightarrow \omega _{A}\text{, }\omega _{\pi
/2\ell }^{(+)}\rightarrow \omega _{C}\text{.}
\end{equation}%
If $\langle \sigma ^{z}\rangle <0$ and $\delta \ll -2g\sqrt{|\langle \sigma
^{z}\rangle |}$, we have
\begin{equation}
A_{\pi /2\ell }\longrightarrow 0\text{, }B_{\pi /2\ell }\rightarrow 1\text{,}
\end{equation}%
and thus
\begin{equation*}
\langle \langle \sigma _{\pi /2\ell }^{-}|\sigma _{\pi /2\ell }^{+}\rangle
\rangle _{\omega }\approx |\langle \sigma _{0}^{z}\rangle |\frac{1}{\omega
-\omega _{\pi /2\ell }^{(+)}}\text{.}
\end{equation*}%
Meanwhile when $g\longrightarrow 0$, the eigen-frequencies
\begin{equation}
\omega _{\pi /2\ell }^{(+)}\rightarrow \omega _{A}\text{, }\omega _{\pi
/2\ell }^{(-)}\rightarrow \omega _{C}\text{.}
\end{equation}

Finally we conclude that if $\delta \gg 2g\sqrt{|\langle \sigma ^{z}\rangle |%
}$, $\omega _{\pi /2\ell }^{(+)}$ is the eigenfrequency of photonic part
while $\omega _{\pi /2\ell }^{(-)}$ is the eigenfrequency of atomic part; On
the other hand if $\delta \ll -2g\sqrt{|\langle \sigma ^{z}\rangle |}$, the $%
\omega _{\pi /2\ell }^{(-)}$ is the eigenfrequency of photonic part and $%
\omega _{\pi /2\ell }^{(+)}$ is the eigenfrequency of atomic part.

\end{document}